\documentclass[10pt,twocolumn]{article}
\usepackage{times}                     
\usepackage{hyperref} 
\usepackage[dvips]{graphicx}
\begin{document}
\title{Rethinking OpenPGP PKI and OpenPGP Public Keyserver}
\author{
 Shinji Yamane\\
 Iwate Prefectural University\\ 
 Faculty of Software and Information Science\\ 
 Takizawa-aza-Sugo, Takizawa vil. Iwate 0200193 JAPAN\\
 {\url{s-yamane@soft.iwate-pu.ac.jp}}
 \and
 Jiahong Wang\\ Iwate Prefectural University\\ 
 pgp-folks@comm.soft.iwate-pu.ac.jp
 \and
 Hironobu Suzuki\\ Independent Software Consultant\\ 
 pgp-folks@comm.soft.iwate-pu.ac.jp
 \and     
 Norihisa Segawa\\ Iwate Prefectural University\\ 
 pgp-folks@comm.soft.iwate-pu.ac.jp
 \and
 Yuko Murayama\\ Iwate Prefectural University\\ 
 Murayama@iwate-pu.ac.jp
}
\maketitle
\thispagestyle{empty}

\begin{abstract}
 OpenPGP, an IETF Proposed Standard based on
PGP\textregistered\ application, has its own Public
Key Infrastructure (PKI) architecture which is different from the one
based on X.509, another standard from ITU. This paper describes
the OpenPGP PKI; the historical perspective as well as its current use.
We also compare three PKI technologies standardized by IETF:
OpenPGP, PKIX(X.509), and SPKI/SDSI.

Since the OpenPGP PKI works without a registration authority nor
certification authority, it fits well with the Internet communication
with voluntary community.
 For example, the digital signature for email including the security
patch program of free software is usually signed by not an authorized
organization but the cross-PGP-signed individuals who belong to
different organizations or nations.

The current OpenPGP PKI issues include the capability of a PGP keyserver and
its performance. PGP keyservers have been developed and operated
by volunteers since the 1990s. The keyservers distribute, merge, and
expire the OpenPGP public keys.
Major keyserver managers from several countries have built the globally
distributed network of PGP keyservers. However, the current PGP Public
Keyserver (pksd) has some limitations. It does not support fully the
 OpenPGP format so that it is neither expandable nor flexible, without
any cluster technology.

Finally we introduce the project on the next generation OpenPGP 
public keyserver called the OpenPKSD, lead by Hironobu Suzuki, one of the
authors, and funded by Japanese Information-technology Promotion Agency(IPA).
\end{abstract}
\setcounter{tocdepth}{4}

\section{Introduction}

Authentication is an essential factor of information security in network
society.
The difficulty of building a Public-Key Infrastructure (PKI) is a major
impediment to strong authentication.
Without PKI, we cannot trust neither digital signature nor certification
based on the public key cryptosystem via wide-area network.

In following section \ref{sec:PKI} and \ref{sec:PKIs}, we overview the
PKI architecture comparing several models. In
section \ref{web_of_trust}, we examine a PKI without authorities which
is presented by the OpenPGP technology and compare it with the other
models.
Then we point out the role of the PGP
keyserver in section \ref{sec:keyserver}
and introduce the next generation OpenPGP public keyserver
project in section \ref{OpenPKSD}. 
Finally we give some conclusions.

\section{PKI architectures}\label{sec:PKI}
PKI has three core functions as follows to manage 
the users' certification and trust relations
~\cite[s.v. ``public-key infrastructure'']{FYI36}:
\begin{enumerate}
 \item to register users and issue their public-key certificates
 \item to revoke certificates when required
 \item to archive data needed to validate certificates at a much later
    time
\end{enumerate}
To operate these three functions with many users on a large-scale network,
many PKI have a hierarchical structure for CAs
and are built using a centralized architecture.
However there are alternatives.

PKI is categorized by the architecture types as follows:
1) hierarchical PKI, 2) mesh PKI, and 3) trust-file PKI~\cite[s.v. ``trusted certificate'']{FYI36}. 
The difference is the way how they rely on CA (Certification Authority).
A hierarchical PKI has the most significant CA in terms of trust 
at the root of the hierarchy tree.
A mesh PKI has CAs issue cross-certificates to each other.
A trust-file PKI has a local file of public-key
 certificates that the user trusts as starting points
 for certification chain.

For example, popular browsers are distributed with an initial file of
trusted certificates, the starting points for certification paths. 
The initial file is different between among the each PKI
architecture. 
 In a hierarchical PKI, the initial file is the root certificate in a
hierarchical PKI. It is usually ``baked into'' the browsers with no 
decisions by the users to trust them. In a mesh PKI, it is the
certificate of the CA that issued the user's own certificate. 
And in a trust-file PKI, any
certificates including self-signed certificates
accepted by the user can be the first public key in a certification path.

\section{PKI standards}\label{sec:PKIs}
To build PKIs, different standards are developed.
They are based on their own framework and architecture and they are never
the same.
This section compares different PKI architectures:
1) X.509 standard from ITU,
2) OpenPGP, an IETF Proposed standard based on PGP\textregistered\
application,
and
3) SPKI, another standards based on the theoretical research.

X.509 is the earliest framework
to provide and support authentication 
including formats for X.509 public-key
certificates, X.509 attribute certificates, and X.509 CRLs.
X.509 is the hierarchical PKI that a CA, central digital certificates
issuer, is responsible for managing the certificates.

Historically, X.509 was standardized by ITU-T (Inter\-national
Tele\-communication Union Tele\-communication sector, formerly CCITT) and
turned to be ISO standard.
X.509 follows the X.500 directory service and provides an example of
reliable authentication and certification.
In practice, developers relax the strict X.500 service scheme.
For example, 
X.509v3 (Version 3) certificate has ``extensions'' field for 
flexible operation.
%

IETF had discussed about the design based on
X.509 framework from each applications to general PKI.
%
Internet standards for X.509 PKI framework is developed 
at IETF Public-Key Infrastructure (X.509) Working Group. 
PKIX not only profiles X.509 standards, but also
develops new standards apropos to the use of X.509-based PKIs in the
Internet.

One of the most popular implementations of X.509-based PKI
is OpenSSL ({\url{http://www.openssl.org/}}, formerly SSLeay).
OpenSSL is a set of Open Source
cryptography libtaries including X.509 CA operation scripts
and distributed freely, such as a part
of PKI package for either commercial or non-commercial purpose.
%


 OpenPGP is the standard based on Pretty Good Privacy\textregistered\
(PGP\textregistered) application which is developed by 
Philip Zimmermann~\cite{garfinkel94}.
PGP\textregistered\ is provided as 
commercial version and `freeware' version for
non-commercial/non-governmental purposes only.

The specification of PGP is standardized as OpenPGP by IETF OpenPGP Working Group.
Today 
``Open\-PGP Message Format'' is defined in RFC2440~\cite{RFC2440}
and to be updated~\cite{RFC2440bis08}.
%
%
The most popular OpenPGP implemenation is 
GnuPG (GNU Privacy Guard), developed by Free Software Foundation
and maintained by Werner Koch of GUUG (German Unix Users Group).
%
%
Either PGP or GnuPG has been known as email cryptography software firstly, 
however they has become the general purpose data encryption tool,
including key exchange over Internet, trust computation, etc.
In the following sections, we examine the only PKI part of OpenPGP.
%

It is worth to point out another possible architecture, as 
we sometimes take an closed binary question such as ``X.509 or OpenPGP,
which is better?'', not as 
``Which PKI will be the appropriate solution for different usage-scenarios?''.
There exists another PKI standardized by IETF --- SPKI (Simple Public Key
Infrastructure).
IETF SPKI Working Group
finished its initial standardization process
and bring into the inter-operation stage~\cite{Ninghui2000,RFC2693}.
It is also called SPKI/SDSI as it is a joint force with SDSI (Simple
Distributed Security Infrastructure) research.
SPKI is designed with distributed and scalable architecture in
many aspects, i.e., no single root CA, no globally distinguished name,
and flexible validity periods~\cite{Adams1991,Adam1997}.


Table \ref{table:SPKI/SDSI} shows the technical comparison 
of X.509, OpenPGP, and SPKI/SDSI
based on the analysis by Clarke~\cite{Clarke2001}.


\section{Certification without Authority}\label{web_of_trust}

\subsection{From Face-to-Face to Web of Trust}
Without a certification authority,
the problem of trusting keys arise 
to assess applicants before giving out certificates.
In OpenPGP, there are no official mechanism for creating certificates,
no officail channel for acquiring and distributing. 
It makes the
process of certification into the face-to-face, {\it ad hoc} situation.
Each end user is respobsible to decide which certificate (public key of
an user) is trusted and accepted to be added into their local trust-file
(denoted ``keyring'' in PGP).

This certification process does not require a trusted, monitored
registration authority or certification authority, however, it lacks
scalability.
So since PGP\textregistered\ 2.0~\cite[pp.~201--203]{Levy2001},
``web of trust'' model that PGP signer acts as an introducer between
people had been supported.

\begin{figure}[h]
 \begin{center}
 \includegraphics[width=3.0in]{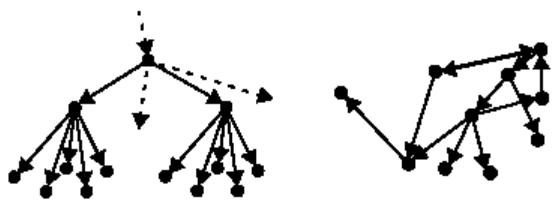}
 \end{center}
 \caption{Hierarchical PKI and Web of Trust~\cite{Caronni2000}}\label{fig:pki.eps}
\end{figure}

Figure \ref{fig:pki.eps} illustrates the model of hierarchical PKI and 
web of trust.

\vspace{1.0cm}




\subsection{Internet Usage Scenarios}
 OpenPGP has its own market which is different with X.509,
and OpenPGP community has grown in a global Internet.

The most famous and critical use might be security alerts.
FIRST (Forum of Incident Response and Security Teams) and its members
including CERT(Computer Emergency ResponseTeam)/CC(Coordination Center)
have their official PGP/GnuPG public keys publicly available~\cite{First2001},
and have signed their alerts with their own PGP/GnuPG key.

Usenet, operated by volunteer NetNews managers, is another example of
the distributed network with OpenPGP PKI. The digital signature for
Usenet control commands should be signed with PGP keys of represented
voluntary managers since 1990s~\cite{pgpcontrol}.

\subsection{What is the Web of Trust?}
\begin{figure}[h]
 \begin{center}
 \includegraphics[width=3.3in]{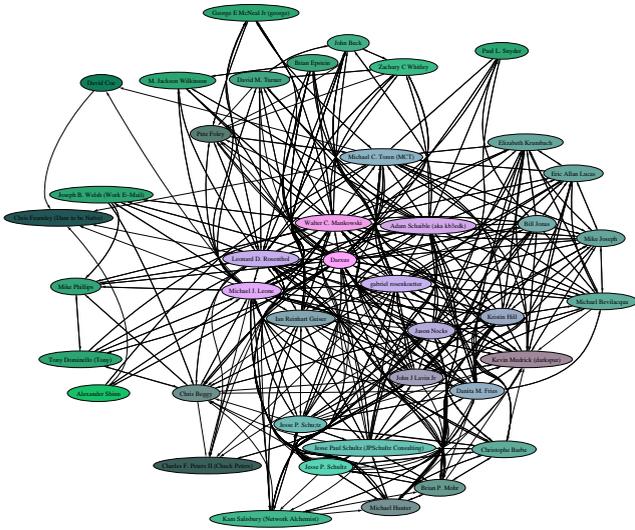}
 \end{center}
 \caption{Visialized Web of Certificate~\cite{springgraph}}\label{fig:phillylinux.ps}
\end{figure}

 OpenPGP provides key management and certificate services
using local trust-file PKI.
The more signature is accepted, the more trust-file generated.
Figure \ref{fig:phillylinux.ps} is an example of a trust-file
visualized the relashonship of OpenPGP signature.
This graph illustrates who introduce the other or who meets with face-to-face,
in other words, whose key signed the others' public key.
There are no central authorities but multiplexed indivisual
relationships in a community.


%

However, this graph is not a ``web of trust'' but just a ``web of
certificates''.
 OpenPGP separate the trustworthy from validity of cerfiticate.
For example, the amount of trust of the introducer and unknown newcomer is
different for an OpenPGP user.
Even if their certification is valid, the issuer of the key is not 
an authority but the users own. 
So OpenPGP users should be responsible on
``Whose keys should be taken as valid but untrusted?''~\cite[p.~81]{Stajano2002}.

In OpenPGP, users issue their signature to the other's public key with
their degree of trust. This is denoted ``trust signature'' and
represented as (trust level, trust amount)
with OpenPGP key management and certificate service.
An ordinary valid signed key is trust level 0, and 
The signed key is asserted to be a valid trusted introducer
is level 1.
Level 2 means ``meta introducer'' or ``introducer-of-introducer''
that its signed key is asserted to be trusted 
to issue level 1 trust signatures.
(Generally, as the introducer is more trustworthy,
a level $n$ trust signature asserts that a key is trusted to
issue level $n-1$ trust signatures.)
The trust amount is in a range from 0--255,
and appointed 60 for partial trust and 120 for complete trust~\cite[s.v. ``Trust Signature'']{RFC2440}.
As OpenPGP distinguished the trust from validity, 
``web of trust'' is also distinguished from ``web of certificates''.


\subsection{PGP Revocation Problem}\label{revocation}

The weakest link of OpenPGP PKI is the revocation of public
key~\cite[pp.~585--586]{Schneier1995}~\cite[p.~309]{P2P2001:ch16}.
As there is no official channel for acquiring and distributing OpenPGP 
public keys, there are no guarantee about 
how to tell everyone that your key is no longer valid.

The typical answer to this revocation problem of PGP is to use
PGP public keyserver for distributing certification.
``Typically, to communicate that a certificate has been revoked, a signed
note, called a key revocation certificate, is posted on PGP certificate
servers, and widely distributed to people who have the key on their
public keyrings. People wishing to communicate with the affected user,
or use the affected key to authenticate other keys, are warned about the
hazards of using that public key''~\cite[pp.~56--57]{Clarke2001}.
However, there are few research on the PGP public keyserver and 
usually the keyserver is not considered as the part of OpenPGP PKI.
In the following, this paper examines PGP keyserver
as the part of OpenPGP PKI.

\section{Related Works}\label{sec:related_works}
There are several research fields related to OpenPGP public keyserver.
The first is the study on the traditional PGP public keyservers,
the second is the integrated channel for OpenPGP key distribution,
and the third is the combined ``web of trust'' with other PKI.

A ``web of trust'' used in PGP is referred in several 
researches including the peer-to-peer authentication~\cite{P2P2001:ch16},
trust computation~\cite{Maurer96b,Caronni2000},
and privacy enhanced technology~\cite{garfinkel94}.
However, there are few description on PGP keyserver.
It might be because PGP keyserver mechanism is too simple.
It is not a CA but just a pool of public keys.
From users' viewpoint, PGP keyserver has a 
large amount of OpenPGP public keys that provide
the interesting material for social analysis of network community. 
For example, OpenPGP keyserver developer Jonathan McDowell
also developed ``Experimental PGP key path finder''~\cite{McDowell2002}
that searches and displays the chain of certification between the users.

As OpenPGP's initial trust file is blank, the users have to start with a
face-to-face certificate to exchange public keys.
Though another initial file is provided via high integrity channel.
{\it Global Internet Trust Register}~\cite{Register}
is a printed book that contains ``fingerprints'' (hash values of certificate)
of the most important public keys (mainly cryptography experts who are
likely to have signed many other keys in their respectice
communities)~\cite[pp.~80--81]{Stajano2002}.

 OpenPGP PKI itself can be described as the superset of PKI~\cite{Zimmermann2001},
however, combining OpenPGP PKI with other authentication system is 
challenging work in both theoretical and operational field.
Formal study of trust relationship of PKI started in the late 1990s~\cite{Maurer96b,Caronni2000} and GnuPG development version in December 2002
started to support its trust calculation with
GnuPGP's trust signature.

The implementation of trust calculation is ongoing and 
using large-scale ``web of trust'' (not ``web of certificates'')
is not so popular outside of computer experts.
On the other hands, using different types of PKI has become more popular.

In the early work at MIT,
PGP-signing service had been combined with Kerberos CA
system that does not have public key cryptography~\cite{Schiller1995}.
Today, the hybrid system of OpenPGP and X.509 is 
both developed into some OpenPGP implementations.
In 2001, German authorities 
BSI (Bundesamt f\"{u}r Sicherheit in der Informationstechnik,
Germany's agency for information technology security)
accept the \"{A}gypten project for Open Source implementation
of governmental mail user agents Sphinx 
which supports X.509v3, PKCS, LDAP, and OpenPGP ~\cite{Newsforge20011005}.
The results of the open development are begun to
imported to other commercial products in 2002--2003~\cite{Kmail}.
In a same way, PGP Corporation also released PGP\textregistered\ version
 8.0 as X.509-enabled application that can interoperate X.509
 certificates and CAs~\cite{PGP8X509}.



\section{OpenPGP Public Keyserver}\label{sec:keyserver}
Before describing our research, this section describes 
OpenPGP keyserver generally.
Keyserver is not a CA. It only pool anyone's public keys.
Keyserver never issue any certificate for someone's public key
but only provide it.

\subsection{Current Status}
The first keyserver is developed at MIT in 1994
by Brian~ A.~LaMacchia. It exchange public keys with email 
and keys are managemented with PGP command.
For users, keyserver acts as an easy email agent
who receives any valid but untrusted keys,
then searches and provides the key to everyone.
%
%
%
In 1997, PGP Public KeyServer (pksd, 
{\url{http://www.mit.edu/people/marc/pks/}}) started 
by MIT student Marc Horowitz. 
Pksd uses a database management system and has been working fine.
The database system is operated via email, CGI-interface from http server,
and HKP --- pksd's own communication protocol over
Hypertext Transfer Protocol (HTTP).
In 2003, David Shaw of GnuPG team proposed the OpenPGP HTTP Keyserver
Protocol~\cite{HKP200303} based on traditional HKP
as the draft for Internet Standard.

Today pksd has been working fine even if in global distributed
network. There are 10 or more syncronized public keyservers in the world
and the most of them are running with patched pksd.
These public keyservers are operated by voluntary managers belong to
organizations including 

MIT

and 

Georgia Tech

in United States,

SURFnet 

in Netherlands,

DFN-CERT

in Germany,

RedIRIS (IRIS-CERT)

in Spain,

JPNIC

in Japan.
Today they have more than 1,400,000 public keys entries and
3,000/day or more transactions between each sync sites.
In 2000,
SURFnet held the first PGP keyserver manager symposium~\cite{SURFnet2000}
and the managers keep in touch with each other. 

\subsection{Revocation process and Keyserver}
As public keyservers provides semi-official key distribution channel,
keyserver adds powerful feature to OpenPGP PKI.
%
Public keyservers can handle the PGP revocation problem that we described in
section \ref{revocation}.
%
%
%
 Using keyserver provides an answer to the question 
``How do you tell everyone that your key is no longer valid?''.
User may issue a suicide note (denoted as ``revocation signature'' in OpenPGP )
and post it to keyservers.
Receiving a valid revocation signature, keyserver updates the key to be
revoked. The update key with revocation signature is redistributed to
the synchronized keyservers in the world, 
and finally PGP user updates their keyrings with the nearest keyserver.
The updated key with valid revoked signature makes users's older key not
to be used.



\subsection{Current Keyserver Problem}
Today's sisutation around PGP keyserver
is beyond the original developers' idea, and 
current pksd also has some limitations.

Firstly, the implementations of pksd are not OpenPGP-compliant.
 OpenPGP ~\cite{RFC2440} published in 1998
defines two versions of signature formats.
(Version 3 provides basic OpenPGP signature information, while version 4 provides
an expandable format with subpackets.)
These changes made traditional PGP applications 
not-OpenPGP-compliant --- not only
PGP\textregistered~\cite[s.v. ``Implementation Nits'']{RFC2440} 
but also pksd.
Today pksd does not fully support OpenPGP format.

Seconary, the pksd does not scalability for global use.
Though pksd has simple but strong dabatabase management system,
it is neither soshisticated nor scalable compared with 
today's Internet server.
For the matter, pksd cannot handle 1 billion keys and cannot
accept such many transactions as Yahoo! or eBay site.
New design of OpenPGP public keyserver is required.

\section{OpenPKSD: Next Generation OpenPGP Public
 Keyserver}\label{OpenPKSD}
We introduce our next generation OpenPGP Public Keyserver project
with a new architecture. We call it OpenPKSD
( OpenPGP Public KeyServer Daemon).
It is developed by one of the
authors and funded by Japanese Information-technology Promotion Agency
 (IPA) in 2001--2002.

\subsection{Server Design and Implementation}
OpenPKSD supports OpenPGP subpacket format
and works as high-performance server with SQL backend.
The design of OpenPKSD oriented to not only high-performance, 
but also flexible extension capability and easy operation.
We implemented OpenPKSD with Ruby and 
PostgreSQL backend~\cite{RubyConf2002}.

\subsection{User Interface and Security}
As ``Web of trust'' depends on users' decision,
user interface is also important factor on security.
For example, Whitten and Tygar~\cite{Whitten1999} had 
ever pointed out some dangerous errors occured with past PGP clients'
interface.

Users can connect to OpenPKSD with two kind of 
interfaces, OpenPGP client or CGI on WWW.
Providing WWW interface, OpenPKSD
must help users' recognization, judgement, and 
handling on OpenPGP public keys.



OpenPKSD displays only 64bit KeyID to identify someone's public keys.
Though some other servers calculate and display ``fingerprint'' of
public keys before download it, it does not help users arare of risk
using keyserver.
As keyserver is just a pool and not CA, 
users should check the public key with their own.
Moreover, it is easy to make some faked keyserver by
Man-in-the-Middle Attack, TCP hijacking, etc.
It means that 
the fingerprint must be calculated under user's (safe) machine
and that is the reason why OpenPKSD does not display fingerprint.

OpenPKSD WWW interface provides additional feature 
to visualize subpackets of PGP keys.
As OpenPKSD has an expandable format with subpackets,
it is very hard to understand the data structure inside this. 
Using pgpdump program,
key packet visualizer that displays the packet format of
 OpenPGP and PGP\textregistered\ version 2.

Many PGP users are familiar with this verification
on added keys, as in 2000, PGP\textregistered\ version 5.5.x to 6.5.3 had
a serious security hole that cannot detected with fingerprint
verification.
Then CERT/CC had alarted 
``Check certificates for ADKs [Additional Decryption Keys]
 before adding them to a keyring.''~\cite{CERT:CA-2000-18}
Pgpdump exactly visualizes these additional keys.
With pgpdump, OpenPKSD helps users to recognize the information of 
public key and any other added keys before downloading.

\subsection{Performance and Future work}
OpenPKSD is implemented with Ruby language and PostgreSQL DBMS.
Ruby is so-called ``scripting language'' and seemed not suitable 
for a quick response or large program development.
However, 
OpenPKSD succeeds not only the more compact code size but also
quick response compared with pksd,
by loading bit calculation modules such as CRC24 checksum
written by C language~\cite{Hironobu2003}.
Table \ref{table:OpenPKSD} shows the performance of 
OpenPKSD version 0.2.8, non-cluster version, installed on PC.
\begin{table}[htbp]
\begin{center}
\begin{tabular}{|c|c|}
 \hline
 CPU: & Intel Pentium4 1.6GHz \\
 \hline
 HDD: & IDE ATA100 7200rpm 60GB\\
 \hline
 Memory: & PC2100 768MB \\
 \hline
 One key query: & 120ms average,\\
 & 72ms best,\\
 & 230ms worst.\\
 \hline
\end{tabular}
 \caption{OpenPKSD Performance}\label{table:OpenPKSD}
\end{center} 
\end{table}
%
%
%
OpenPKSD version 0.2.8 is also working 
well at handling usual transaction between other PGP public keyserver
described in section \ref{sec:keyserver} since 2002.

Forthcoming developers' version of OpenPKSD will support some clustering
based on the reserch on the performance of 
cluster technology~\cite{Wang2002:ICPADS}.
It will be published in 2003 and 
support the experimental HKP(keyserver protocol over http) balancer,
keyserver cluster, and clustered database.

\section{Summary}
In this paper, we overlooked some PKI architectures.
%
Using ``Web of Trust,'' OpenPGP PKI can help users to
manage certification without CAs.
However, there are the problem on public key management, i.e., 
how to get the receivers' public key, or, 
how to tell everyone that the public key is no longer valid.
PGP keyserver is the solution to the problem.

Though some PGP public keyservers have built a global PKI,
traditional PGP keservers have some limitations.
We introduced OpenPKSD, 
newly-designed and OpenPGP-supported 
public keyserver project.
OpenPKSD took its first step, works well in practice, and 
examining the cluster technology.

\section*{Availability}
OpenPKSD source code and documents
are available under GNU General Public License (GPL) at 
{\url{http://www.openpksd.org/}}.


\onecolumn
\begin{table}[htbp]
 \begin{center}
{\small 
 \begin{tabular}{|l|r|l|}
  \hline
{\bf X.509}  & CA Characteristics: & Global Hierarchy. There are commercial X.509 CAs. \\ 
  & & X.509 communities are built from the top-down. \\ 
  & Trust Model: 
  & Hierarchical Trust Model. Trust originates from a `trusted' \\ 
  & & CA, over which the guardian may or may not have control. \\ 
  & & A requestor provides a chain of authentication from the \\ 
  & & `trusted' CA to the requestor's key. \\ 
  & Signatures: & Each certificate has one signature, belonging to the issuer \\ 
  & & of the certificate. \\ 
  & Certificate Revocation: & Uses CRL(Certificate Revocation List)s \\ 
  & Name Space: & Global \\ 
  & Types of Certificates: & Name Certificates \\ 
  & Name-to-Key binding:
  & Single-valued function: each global name is bound to ex- \\ 
  & & actly one key (assuming each user has a single public- \\ 
  & & private key pair). \\ 
  \hline
 {\bf OpenPGP} & CA Characteristics: & Egalitarian design. Each key can issue certificates. \\ 
 \quad {\bf } & & PGP communities are built from the bottom-up in a\\ 
  & & distributed manner. \\ 
  & Trust Model: & {\it Web of Trust}, file-based PKI.\\ 
  & Signatures: & Each certificate can have multiple signatures; the first \\ 
  & & signature belongs to the issuer of the certificate. \\ 
  & Certificate Revocation:  & A `key revocation certificate,' suicide note is posted on \\ 
  & & {\it PGP keyservers}, and widely distributed to people who\\ 
  & & have the compromised key on their public keyrings. \\ 
  & Name Space: & Global \\ 
  & Types of Certificates: & Name Certificates \\ 
  & Name-to-Key binding: 
  & Single-valued function: each global name is bound to ex- \\ 
  & & actly one key (assuming each user has a single public-\\ 
  & & private key pair). \\ 
  \hline
 {\bf SPKI/~SDSI} & CA Characteristics:  & Egalitarian design. The principals are the public keys. \\ 
 {\bf } & & Each key can issue certificates. SPKI/SDSI communities \\ 
  & & are built from the bottom-up in a distributed manner. \\ 
  & Trust Model:  & Trust originates from the guardian. A requestor provides \\ 
  & & a chain of authorization from the guardian to the \\ 
  & & requestor's key. The infrastructure has a clean, scalable \\ 
  & & model for defining groups and delegating authority. \\ 
  & Signatures: & Each certificate has one signature, belonging to the
  issuer\\ 
  & & of the certificate.\\ 
  & Certificate Revocation: & Advocates using short validity periods
  and {\it Certificates of}\\ 
  & & {\it Health}.\\ 
  & Name Space: & Local \\ 
  & Types of Certificates: & Name Certificates,
  Authorization Certificates \\ 
  & Name-to-Key binding:  & Multi-valued function: each local name is bound to zero, \\ 
  & & one or more keys (assuming each user has a single public \\ 
  & & -private key pair). \\ 
  \hline
 \end{tabular}}
 \end{center}
 \caption{Comparison of X.509, OpenPGP, and SPKI/SDSI}\label{table:SPKI/SDSI}
\end{table}
\twocolumn


\end{document}